\documentclass[sigconf,authorversion,nonacm]{acmart}

\AtBeginDocument{%
  \providecommand\BibTeX{{%
    \normalfont B\kern-0.5em{\scshape i\kern-0.25em b}\kern-0.8em\TeX}}}

\usepackage{subcaption}

\begin{document}

\title{Hybrid moderation in the newsroom: Recommending featured posts to content moderators}

\author{Cedric Waterschoot}
\orcid{1234-5678-9012}
\email{cedric.waterschoot@meertens.knaw.nl}
\affiliation{%
  \institution{KNAW Meertens Instituut}
  \streetaddress{Oudezijds Achterburgwal 185}
  \city{Amsterdam}
  \country{The Netherlands}
  \postcode{1012 DK}
}

\author{Antal van den Bosch}
\affiliation{%
  \institution{Institute for Language Sciences, Utrecht University}
  \city{Utrecht}
  \country{The Netherlands}}
\email{a.p.j.vandenbosch@uu.nl}

\begin{abstract}
Online news outlets are grappling with the moderation of user-generated content within their comment section. We present a recommender system based on ranking class probabilities to support and empower the moderator in choosing featured posts, a time-consuming task. By combining user and textual content features we obtain an optimal classification F1-score of $0.44$ on the test set. Furthermore, we observe an optimum mean NDCG@5 of $0.87$ on a large set of validation articles. As an expert evaluation, content moderators assessed the output of a random selection of articles by choosing comments to feature based on the recommendations, which resulted in a NDCG score of $0.83$. We conclude that first, adding text features yields the best score and second, while choosing featured content remains somewhat subjective, content moderators found suitable comments in all but one evaluated recommendations. We end the paper by analyzing our best-performing model, a step towards transparency and explainability in hybrid content moderation. 

\end{abstract}



\maketitle

\section{Introduction}
Online newspapers allowing user comments have been facing moderation challenges with large and increasing content streams \citep{Meier2018,Wintterlin2020}. Whether to filter out toxicity, counter misinformation, or promote constructive posts, platforms are looking towards computational solutions to support moderator decisions \citep{Gollatz2018}. Overall, moderation strategies are focused on two polar opposites. On the one hand, the moderator is required to safeguard the comment space from toxic and negative content \citep{Wintterlin2020}. On the other hand, platforms aim to promote what they deem \textit{good} contributions, for example by pinning certain content to the top of the page \citep{Diakopoulos2015}.

In this paper we present a recommender based on ranking class probabilities to support the content moderator in picking such featured posts. Using Dutch comment data with human labeling of featured posts, we train a set of models which present the human moderator with a set of posts that might qualify for being featured. We hypothesize that the optimal post representation for ranking includes both user features and textual content features, information used by content moderators as well. Furthermore, we validate our models separately on a collection of articles. Validation on unseen articles reflects the real-life setting of moderating and choosing only a few comments to be featured, as opposed to artificially split and balanced test sets. The output of the best-performing model is assessed in an expert evaluation by content moderators currently employed at the platform in question, who evaluated a random selection of articles by deciding whether the recommended comments are worthy of getting featured. 

\section{Background}
\subsection{Online Content Moderation}
As online comment platforms grow, content moderators have had to adapt moderation strategies to the changing online environments. Dealing with negative content has been a particular focus, e.g. detecting trolling and online harassment \citep{quandt2022} or even organized misinformation campaigns \citep{Meier2018}. \citet{Quandt2018} describes these forms of negative content under the umbrella of 'dark participation'.
Recently, however, moderators are seeking to promote \textit{good} comments as well. On the opposite side of the comment spectrum from dark participation, platforms and moderators are selecting what they deem as good, high-quality comments and manually pinning them to the top of the comment space.  

Promoting what news outlets see as high-quality contributions has for example taken the form of New York Times (NYT) picks \citep{Diakopoulos2015}, Guardian Picks at The Guardian, or featured posts at Dutch news outlet NU.nl. On their FAQ pages, these outlets describe such promotion-worthy comments as "substantiated", "respectful" and representing "a range of viewpoints"\footnotemark. \citet{Diakopoulos2015} assigns a set of twelve editorial criteria to such featured posts, ranging from argumentative quality to entertainment value and relevance. Overall this procedure may be seen as "a norm-setting stategy"  (\citealp[p.4]{Wang2022}). The authors argue that exposure to these promoted posts may also improves the quality of succeeding comments \citep{Wang2022}. 

\footnotetext{\url{https://www.nu.nl/nujij/5215910/nujij-veelgestelde-vragen.html}}
\footnotetext{\url{https://help.nytimes.com/hc/en-us/articles/115014792387-The-Comments-Section}}

Supplementary to the aforementioned goal of promoting high quality content and the positive normative effects these posts may have on other commenters, user engagement may increase as well. \citet{Wang2022} find that after a user received their first featured comment, their commenting frequency increased.

\subsection{Hybrid Moderation}
While featured content ranking for moderators is a novel task, recommender systems have been used in the context of news platforms. Plenty of research and workshops (e.g. INRA workshops) focus on news recommendation and personalization aimed at readers on these platforms \citep{Raza2022}. While this application is adjacent to content moderation, it differs from our application in that it is mostly aimed at users of a platform (as opposed to moderators) to optimize news consumption (instead of improving moderation tasks). 

Moderators of online news outlets have been increasingly working with computational systems to perform their tasks to ward off toxic and unwanted content \citep{Delort2011,Gorwa2020}. The result is a hybrid setting in which the role of human moderator on the one hand, and the computational system on the other hand have been intertwined. \citet{Ruckenstein2020} argue that ideally, AI should offer decision support to the human moderator. Taking the final decision on publishing content is a task exclusively for the human moderator, and tools should be focused on assisting this function \citep{Ruckenstein2020}. \citet{Park2016} emphasize this hybrid relation, stipulating that journalists do not want automatic editorial decision-making. When computational moderation tools support the human in carrying out their tasks, the moderator themselves can adapt to the nuances of changing online contexts and apply human interpretation and judgement \citep{Park2016}.

Classifying toxic comments in online comment spaces has received substantial attention \citep{Gorwa2020,Wang2021}. The classification of featured comments or editor picks, however, has not been explored quite that often. \citet{Diakopoulos2015b} uses cosine similarity to calculate relevance scores relative to the conversation and the article using New York Times editor picks. The author discovers an association between these picks and relevance and concludes that such computational assistance may speed up comment curation \citep{Diakopoulos2015b}.
As part of their CommentIQ interface, \citet{Park2016} train a SVM classifier on unbalanced, but limited, data ($94$ NYT picks, $1,574$ non-picks) and achieve a precision score of $0.13$ and recall of $0.6$. Their data includes user history criteria as well as comment variables \citep{Park2016}.

\citet{napoles-etal-2017-finding} annotated comments from Yahoo News in terms of what they present as "ERICs: Engaging, Respectful, and/or Informative Conversations". The authors look at the constructiveness of a thread rather than a single comment, and do not use editorial choices as their labelling \citep{napoles-etal-2017-finding}. 
\citet{kolhatkar-taboada-2017-using} combined the Yahoo comments with NYT picks. The authors achieve an F1-score of $0.81$ by training a BiLSTM on GloVe embeddings, using the NYT picks as benchmark and a balanced test set \citep{kolhatkar-taboada-2017-using}. Furthermore, they combine a set of variables, including comment length features and named entities, and achieve a best F1-score of $0.84$ using SVMs \citep{kolhatkar-taboada-2017-using}. In a follow-up study, the authors achieved an F1-score of $0.87$ on a similar task using crowdsourced annotations and logistic regression \citep{kolhatkar2020}.

To sum up, these classifiers for the most part lacked input aside from comment information, whether text representation or otherwise. Additionally, the validation of these models was performed on large, balanced test sets, which does not resemble the real-life practice of picking featured posts. The moderator chooses editor picks on the article level and any model should therefore be evaluated on such tasks. In this paper, we combine user information with comment data and text representation, all information used by the moderators themselves.

\subsection{Platform Specifics}

The comment platform discussed in this paper is called NUjij, part of the Dutch online newspaper NU.nl\footnote{\url{https://nu.nl/}}. NUjij allows commenting on a wide range of pre-selected news articles. Pre-moderation safeguards the comment space, in the form of automatic toxicity filtering and human moderators checking the uncertain comments \citep{VanHoek2020}. NUjij employs a selection of moderation strategies, including awarding expert labels to verified users and presenting featured comments above the comment section, similar to New York Times or the Guardian picks \citep{NU.nl2020}. Picking featured comments is done by human moderators and are described as "substantiated and respectful reactions that contribute to constructive discussion"\footnote{\url{https://www.nu.nl/nujij/5215910/nujij-veelgestelde-vragen.html}}. Nujij states in their FAQ that moderators aim to present balanced selections and to not pick based on political affiliations. This paper aims to address this specific task making use of the information available to moderators, which includes user information and history. Other platforms might have different editorial guidelines for moderators to choose featured comments. To best support the moderator, it is important that the approach fully suits their context, which may include the (intended) human bias in picking featured content.

\section{Methodology}

\subsection{Data}
We obtained a Dutch language dataset containing a total of $821,408$ pseudonymized posts from the year 2020, spanning $2,952$ articles from NU.nl\footnotemark. Major topics within this dataset are climate change, the 2020 US election and the COVID-19 pandemic. A binary variable indicates whether each post was featured by a moderator during the time interval that commenting on the page was allowed. User variables were obtained by grouping and aggregating the information across the pseudonymization user keys. In total we have $8,262$ featured posts. An article has on average $2.8$ featured posts (sd=$3$), with a median of $2$. The average article has $278$ comments (sd=$358$, median=$173$). This shows that, while large variation exists in the number of comments per article, the number of featured posts per article remains low and relatively stable. The number of featured posts does not grow along with the number of comments posted per article and, therefore, articles with many comments cause great difficulty in finding the featured posts. 

\footnotetext{\url{https://anonymous.4open.science/r/HybridModeration_RecSys2023/README.md}}

\begin{figure}[h]
    \centering
        \includegraphics[width=1\columnwidth]{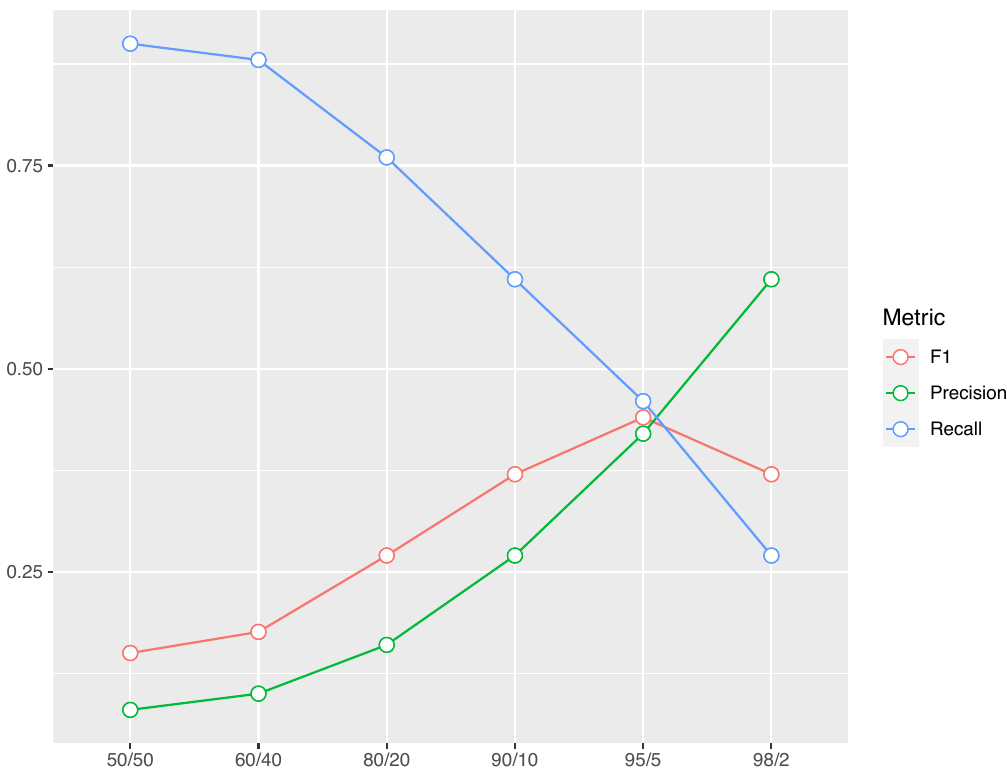}
\caption{Featured post vs non-featured post ratios, metrics calculated on validation data (n=$32,122$)}
\label{splits2} 
\end{figure}

We group the comment data by article\_id and sort these chronologically. We split this data $50\%/50\%$, resulting in two sets of $1,476$ articles, with the split date at June 16th 2020. The first set of articles is used for training and testing classifiers. We further split this set into $80\%/10\%/10\%$ generating a training, validation and test set, respectively. Table \ref{splits} shows the distribution of posts in each set. Using the validation data, we tested the downsampling of the non-featured posts in the training set using all the featured posts in the training data (n=$3,047$). Using the features listed in Table \ref{vars}, we trained a random forest to predict if a post was featured on six different downsampled training sets (Figure \ref{splits2}). The $95/5$ ratio, i.e. 95\% non-featured posts and 5\% featured posts, yielded the best result and will be used as the training data henceforth. While the $95/5$ ratio still remains unbalanced, it is important to note that the unsampled actual ratio approximates $99/1$. Thus, the training data represents a marked downsampling of non-featured posts.

\begin{table}[htb]
        \centering
        \small
        \setlength\tabcolsep{2pt}
\begin{tabular}{  l c c }
  \toprule
\textbf{Dataset} & \textbf{Total Posts} & \textbf{Featured}  \\
  \midrule
Training & 256,973 & 3,047\\
Validation & 32,122 & 384\\
Test & 32,122 & 347\\
95/5 ratio & 60,465 & 3,047\\
  \hline
\end{tabular}
\caption{Article set 1: datasets for training classifiers}
\label{splits}
\end{table}

The second dataset contains $1,476$ articles, published after June 16th 2020, with a total of $500,191$ posts, of which $4,484$ are featured. This second article set is used for evaluating the ranking of unseen discussions.

\begin{table*}[]
        \centering
        \small
        \setlength\tabcolsep{2pt}
\begin{tabular}{  l l }
  \toprule
\textbf{Var name} & \textbf{Description}  \\
  \midrule
Delta\_minutes & Minutes between article publication and comment publication (uptime)\\
Reply\_uptime / reply\_count & Number of replies to the post relative to uptime / absolute number of replies\\
Respect\_uptime / respect\_count & Number of likes relative to uptime / absolute number of likes\\ 
wordcount & Number of words in the comment\\
wordspersentece & Mean sentence length within the comment\\
Total\_posts\_user & Total posts by user\\
Featured\_posts\_user & Total featured posts by user\\ 
Ratio\_featured & Featured posts relative to total posts by user\\
Ratio\_rejected & Rejected posts relative to total posts by user\\
Ratio\_reply & Average reply count on posts by user\\
Ratio\_respect & Average number of likes on posts by user\\
  \hline
\end{tabular}
\caption{Non-textual features}
\label{vars}
\end{table*}

\subsection{Models}
The first model is trained exclusively on the non-textual features listed in Table \ref{vars}. These features are available to the moderator and can be taken into account when deciding to feature a comment. The other two random forest classifiers, detailed later, either include textual Bag-of-Words or word embedding features, while we have also finetuned a transformer-based model on textual input only. Other models were trained (including SVM and logistic regression) but did not perform as well as the random forest implementations.

\subsubsection{Baseline}
We defined a simple threshold-based model to determine whether a post is classified as featured. More specifically, comments posted by users that have a featured post ratio above 3\% will be labelled as such. The threshold constitutes the 95th-percentile. Users with a history of often writing featured comments might do so in a new discussion. To make recommendations, the featured ratio is sorted in descending order.

\subsubsection{Random Forest (RF)}
We trained a random forest on the non-textual variables presented in Table \ref{vars}. We used the standard sci-kit implementation of random forest and performed a hyperparameter grid search\footnotemark. The final model has a max depth of $50$, $200$ estimators and a minumum sample split of $10$. 

\footnotetext{\url{https://scikit-learn.org/stable/index.html}, v1.2.0}

\subsubsection{RobBERT}
Previous models were entirely trained on non-textual data. To obtain a model that uses pure text as input, we employed the pre-trained Dutch transformer-based language model, RobBERT, and finetuned it on our training data \citep{Delobelle2020}. This training data consists of the text belonging to the exact comments the non-textual variables represent. The sequence classification employs a linear classification head on top of the pooled output \citep{Wolf2020,Delobelle2020}. We trained for $10$ epochs with a batch size of $64$, AdamW optimizer and a learning rate of $5e^{-5}$ \citep{Loshchilov2019}. 

\subsubsection{RF\_Emb \& RF\_BoW}
By extracting the embeddings from the previously discussed RobBERT model, we are able to combine the textual input with the set of non-textual variables. We extracted the CLS-tokens from the input and added those to the training data as features, adding up to a total of $797$ features. We trained a sci-kit random forest on this combined data. A hyperparameter grid search was performed leading to a final random forest model with a max depth of $64$,  $1,200$ estimators and min\_sample\_split of $2$. 

Our final model adds another text representation to the mix. Instead of the CLS token used for the RobBERT embeddings, we represented the content by a standard Bag-Of-Words approach, counting the occurrences of the tokens in each comment. First, the text was lowercased and both punctuation and stopwords were removed. The words were added to the non-textual training data, resulting in a set of $426$ features. We once again performed a hyperparameter grid search that resulted in RF\_BoW with $1,200$ estimators, min\_sample\_split of $10$ and a max depth of $110$.

\section{Results}

The initial evaluation is done on the test set that was obtained out of our original 80/10/10 split on the first set of articles. Evaluation on the test set follows the standard procedure of a classification problem, in which comments are not yet ranked by class probability. Next, we use the second set of articles to calculate recommendation scores. In order to recommend comments to the moderator, the model ranks all the posts by class probability. The top comments, i.e. those with the highest probability of belonging to the featured class, are recommended. We evaluate recommendations on the basis of Normalized Discounted Cumulative Gain (NDCG). We calculated NDCG at different recommendation set sizes ($k=3, 5, 10$) across unseen articles. 

\subsection{Classification results}
The initial evaluation step concerned the classifier's generalization performance on the test set, containing a total of $379$ featured posts and $31,743$ non-featured posts. The 'Informed Baseline' achieved an F1-score of $0.17$. This model performed well in terms of recall, but not in precision. The transformer-based RobBERT model, which lacks the non-textual information that the others models have, underperformed as well (Table \ref{metrics}). This might be the case due to the fact that identical comments are sometimes featured and other times not. Due to the limit on featured comments per article, only a small set of well-written comment have the featured label. The RF without textual representation achieved the best F1-score on the test set, while RF\_BoW outperformed the other models in terms of precision (Table \ref{metrics}).

\begin{table}[h]
        \centering
        \small
        \setlength\tabcolsep{2pt}
\begin{tabular}{  l c c c}
  \toprule
\textbf{Model} & \textbf{Precision}& \textbf{Recall}& \textbf{F1-score}  \\
  \midrule
Informed baseline & 0.11& 0.40& 0.17\\
RF & 0.45 & \textbf{0.44} & \textbf{0.44}\\
RobBERT & 0.09& 0.25& 0.13\\
RF\_emb &0.24 &0.29 & 0.26\\
RF\_BoW & \textbf{0.55} & 0.32 & 0.41 \\
  \hline
\end{tabular}
\caption{Evaluation on test (n=$32,122$ \& $347$ featured)}
\label{metrics}
\end{table}

\subsection{Validation on unseen articles}

Besides the standard classification of rare featured posts, these positively classified comments ought to be part of the recommendation set as well. Ranking is done by sorting individual comments by class probability for the featured class in descending order. A recommendation consists of the top $k$ posts derived from this ranking.

To validate the models, we calculated NDCG@k, with $k$, the number of comments recommended, at $3$, $5$ and $10$ across all $1,476$ articles \citep{Jarvelin2000}. An article has on average $3$ featured posts, while $5$ and $10$ allows for the moderator to choose.

\begin{table}[h]
        \centering
        \small
        \setlength\tabcolsep{2pt}
\begin{tabular}{  l c c c}
  \toprule
\textbf{Model} & \textbf{NDCG@3}& \textbf{NDCG@5}& \textbf{NDCG@10}  \\
  \midrule
Informed baseline & 0.42& 0.47& 0.50\\
RF & 0.86 & 0.86 &0.83\\
RobBERT & 0.33& 0.39&0.45 \\
RF\_emb & 0.52& 0.57& 0.60\\
RF\_BoW & 0.87 & 0.87 & 0.86\\
  \hline
\end{tabular}
\caption{NDCG averaged across $1,476$ unseen articles}
\label{ndcg}
\end{table}

The results are shown in Table \ref{ndcg}. The best performing models on the initial evaluation set also yield the best rankings of unseen comments. RF and RF\_BoW performed the best at all recommendation sizes, with the latter yielding the highest score. This result indicates that added text representation in the form of Bag-of-Words slightly improves the recommendations shown to the moderators (Table \ref{ndcg}). Simply ranking comments based on the featured history scored better than ranking based on content, potentially because there is no consistent featuring of well-written comments.

\subsection{Expert evaluation by moderators}
Using the best performing model, a random set of unseen articles was collected alongside the recommendations. We created a survey consisting of $30$ articles combined with a set of comments. This set consisted of the recommended comments (comments with class probability above $0.5$ and maximum $10$ per article) and an equal number of random non-recommended comments from the discussion. These were randomly shuffled so the moderators did not know which comments were recommended by our system. Along with the article and comments, the evaluation included features that moderators have access to in the real-life practice: the number of previously posted and featured comments by the user, the rejection rate of the user and the respect points of the comment. The content moderators had to decide for each individual comment whether they thought it was a candidate to feature on NUjij. In total, four moderators took the survey and each of them labelled comments from $15$ articles. The first five articles were shown to all moderators in order to calculate inter-annotator agreement, while the other $10$ were randomly selected from the pool. We calculated a Krippendorff's alpha inter-rater agreement of $0.62$. This result, combined with the fact that $42.3$\% of comments featured in the original data were not chosen, indicates that picking featured content remains somewhat subjective. 
However, in all but one article, moderators found comments to feature among the recommendations, resulting in a NDCG score of $0.83$. While there is subjectivity involved in picking featured comments, the moderators do find featured content within the recommendations made by the model. They might not all choose the exact same comments, but all find worthy content in the recommended set.

\section{Discussion}
The context of hybrid moderation asks for insight into computational models employed in the pipeline. Transparency being a key value in the field of journalism, moderators and users alike demand explanations as to how models come to a certain output \citep{Ruckenstein2020,Molina2022}. Transparency is a prerequisite for user trust in content moderation \citep{Brunk2019}. Here, we offer an error analysis of our best performing model. Moderators may use this information to counter potential bias towards certain comment characteristics. Furthermore, we discuss the limitations of our approach.

\begin{table}[h]
        \centering
        \small
        \setlength\tabcolsep{2pt}
\begin{tabular}{  l c c c c}
  \toprule
\textbf{Feature} & \textbf{True pos.}& \textbf{False pos.}& \textbf{True neg.}  & \textbf{False neg.} \\
  \midrule
Respect\_count & 37& 28& 15 & 3\\
Ratio\_featured & 4.3\% & 2.3\% & 0.2\% & 1.5\%\\
Wordcount & 107& 111& 79 & 49\\
  \hline
\end{tabular}
\caption{Analysis of (in)correct predictions: means of highest contributing features}
\label{error}
\end{table}

To explain our model's behavior in general terms, we explore the erroneous recommendations the model has made, more specifically which features repetitively contributed to false positives (FP), and false negatives (FN). For the error analysis, we processed all $1,476$ validation articles and collected the top false positives in each recommendation (at k=$5$) and all false negatives. The latter are gathered from the entire article dataset, since they were incorrectly omitted from the actual recommendation. We used the python library 'treeinterpreter' to collect for each prediction the feature contribution\footnotemark. The contribution ($c$) equals the percentage points (as decimal) the feature has contributed to the class probability of the prediction, calculated by following the decision paths in the trees.

\footnotetext{\url{https://github.com/andosa/treeinterpreter}}

Respect\_count ($c=0.14$) and respect\_uptime ($c=0.11$) contributed highly to incorrect recommendations, indicating that that our model often incorrectly recommended posts with a high number of likes (Table \ref{error}). Additionally, the model is biased towards users who have been often featured before (c=$0.06$), and towards longer posts (c=$0.04$). 

Next, we looked at the false negatives (FN). Similar to FPs, the history of being featured is a crucial factor in incorrectly omitting posts. Posts by users that have not been featured before or have an extremely low ratio of featured posts (c=$0.05$) were missed, as can be seen in Table \ref{error}. Furthermore, featured posts with a noticeably low respect\_count (c=$0.09$) were missed as well. Another source of erroneous rankings was wordcount (c=$0.02$). Featured posts tend to be longer (mean featured = $100$, mean non-featured = $53$). Shorter comments may have been overlooked and omitted from the recommendation. 

\subsection{Limitations \& future research}
We see at least two limitations to our approach. First are those related to the platform. Our models make use of a wide range of variables, including aggregated user information which may not be available for other platforms. Furthermore, our recommendations are based on historical moderation choices and may therefore be biased towards certain content. These choices reflect the editorial interpretation of a \textit{constructive} comment by the platform. Future research could compare different criteria for featuring posts. Another platform-related limitation is the language. All text in this study was Dutch. Although we did not test the approach on data in another language, our approach, which assumes the presence of pre-labeled featured post data and a transformer language model for that language, is language-independent.

Second, while we have validated our models on a large collection of articles which resemble the real-life application, we do not know the precise moment at which the moderator selected featured posts. Knowing which posts were available to the moderator at that point in time would allow us to  replay the recommendation process in time-realistic detail. Future research will specifically address this issue, using time-stamped data that documents the precise moment moderators selected featured posts.

\section{Conclusion}

In this paper, we presented a classifier-based recommender system for featured posts to offer decision support to the online content moderator. Using comment and moderation data from a Dutch news platform, we showed that supplementing the non-textual data with text representation achieves the best ranking scores. More specifically, our random forest supplemented with Bag-Of-Words representations achieved the best ranking. While previous research on classifying constructive comments validated their models only on an artificially balanced test set, we validated our models on a large set of articles, replicating real-life practice. Furthermore, content moderators of the platform in question evaluated the output, yielding a NDCG of $0.83$. We unpacked our best performing model in terms of error analysis, showing that our model favoured posts from users with a history of being featured before and might omit comments with a lower respect count. 

With our proposed and novel approach combined with transparency, we aim to support and empower the online content moderator in their tasks, while not obscuring the nuance and contextuality of picking featured posts.


\begin{acks}
This study is part of the project Better-MODS with project number 410.19.006 of the research programme 'Digital Society - The Informed Citizen' which is financed by the Dutch Research Council (NWO).
\end{acks}

\bibliographystyle{ACM-Reference-Format}
\bibliography{sample-base, custom}

\end{document}